\documentclass[12pt]{article}

\usepackage{graphicx}
\usepackage{epsf}

\begin{document}

%%%%% The following lines create the SLAC Pub Title Page
%%
\thispagestyle{empty}
\renewcommand{\thefootnote}{\fnsymbol{footnote}}

%%%%% Substitute your Pub number, month and year in the following:
%%
\begin{flushright}
{\small
SLAC--PUB--8651\\
BABAR-PROC--00/38\\
October 2000\\}
\end{flushright}

\vspace{.8cm}

%%%%% Title and Author Information:
%%
\begin{center}
{\bf\large   
PIN Photodiodes for Radiation Monitoring and Protection \\ 
in the BABAR Silicon Vertex Tracker\footnote{Work supported by
Department of Energy contract  DE--AC03--76SF00515.}}

\vspace{1cm}

T.I. Meyer\\
Stanford Linear Accelerator Center, Stanford University,
Stanford, CA  94309\\

\medskip

representing the BABAR Collaboration
\end{center}

\vfill

\begin{center}
{\bf\large   
Abstract }
\end{center}

\begin{quote}
We discuss the design, implementation and performance of
the radiation monitoring and protection system used by the Silicon
Vertex Tracker (SVT) in the BaBar detector. Using 12 reverse-biased
PIN photodiodes mounted around the beampipe near the IP, we are able
to provide instantaneous radiation dose rates, absorbed dose
integrals, and active protection that aborts the circulating beams in 
the PEP-II storage ring when radiation levels exceed user-defined
thresholds. The system has reliably protected the SVT from excessive 
radiation damage and has also served as a key diagnostic tool in 
understanding radiation backgrounds at PEP-II.
\end{quote}

\vfill

%%%%%%%%%%%%%%%
%% Choose"Presented at," "Contributed to" for conference papers
%% or "Submitted to" for journal papers
%%%%%%%%%%%%%%%
\begin{center} 
{\it (Contributed to)} \\
{\it The Meeting of } \\
{\it The Division of Particle and Fields } \\
{\it of The American Physical Society} \\
{\it Columbus, Ohio, USA} \\
{\it August 9--August 12, 2000} \\
\end{center}

\newpage
%%
%%%%% End of title page

%%%%% Following are the commands to create the rest of the SLAC Pub.

%%%%% Your paper starts here:
%%

%% To get page numbers in the rest of the paper:
%
\pagestyle{plain}

\section{Motivation}

\noindent
The unprecedented beam currents and luminosities at the  PEP-II accelerator
make machine backgrounds a significant challenge to the BaBar experiment.
Despite significant use of radiation hard technology, the Silicon
Vertex Tracker (SVT) is the most radiation vulnerable sub-detector
because of its proximity to the beamline. The SVT has 
established a rigorous program of radiation monitoring and protection (SVTRAD)
to ensure the proposed lifespan, including the ability to
automatically abort the beams in PEP-II when radiation exceeds
programmable thresholds$^1$.

\section{Design}

\noindent
Reverse-biased large-area PIN photodiodes (Hamamatsu S3590-08) were 
selected to meet the multiple constraints of space, expense, and scalability.
Signal current is generated from the
deposition of energy from incident radiation that creates electron-hole
pairs which flow under the applied field. (200 pC $\sim$ 1 mRad of 
absorbed dose)  By continuously accumulating the
the instantaneous dose rates, absorbed dose integrals can
be formed with well-controlled fractional errors. Precision signal extraction 
depends largely on knowledge of the pedestal, which 
depends on temperature and absorbed dose (through radiation damage).

The SVTRAD system uses 2 rings of 6 radiation sensors at
z=$\pm$ 20 cm from the interaction point at a radius of 3 cm. 
Each photodiode is accompanied by 2 closely neighboring thermistors.
Shielded signal cables carry the diode currents
and bias voltage from the sensors to the front-end electronics module
within the SVT shielding scheme.

\section{Implementation}

\noindent
The SVTRAD module is designed to be autonomous and robust$^2$. It maintains
configuration information between power-on resets and
internally buffers messages for remote client retrieval. Each channel in
the SVTRAD module can be configured to be used either for radiation
monitoring or protection. Channels used in the protection
circuit are passed back to the monitoring path, but resolution is degraded.

Each SVTRAD module handles three channels of current input. The
monitoring circuit uses a set of three
charge-integrating 20-bit Burr-Brown DDC101 ADCs, two sets of memory buffers,
and a Xilinx 4013E FPGA. The ADCs are configured to
average 4 on-chip 0.25 ms conversions to deliver data at 1 kHz. This
data is split into two streams (``slow'' and ``fast'') which are
directed into separate registers. The slow datastream is internally 
accumulated within the FPGA and rolled out to fill the 
8K memory at 0.47 Hz.  The fast memories implement a circular 
history buffer with two back-to-back 32K buffers, which can be read
out at remote request.  The fast history buffer provides an average
history of 5.7 seconds at 1 kHz sampling.

The primary radiation concern for the SVT is integrated dose, and
therefore the radiation protection algorithm focuses on placing a
threshold on a minimum (integrated) dose absorbed over a minimum time.
The algorithm enforces a chronic dose rate threshold
(i.e. mR/s), but with a dose tolerance measured in mR.  The
tolerance parameter sets how much dose is integrated
above the chronic dose rate threshold before the circuit trips. Short
term departures over the chronic dose rate threshold are therefore
allowed, so long as their integral is less than the tolerance.

Implementation uses a set of Analog Devices AD652s (so-called 
current-to-frequency converters) and an Orca 2C26A FPGA. The 
FPGA maintains a multi-input counter, whose inputs are the 
converted diode frequency, an internal programmable timer, and the 
thermistor current converted frequency. The diode and thermistor frequencies
are oppositely signed and the programmable timer rate is chosen to 
enforce the desired threshold. When the count value exceeds the 
(programmable) depth of the counter, the circuit trips. 

\section{Performance}

\noindent
The SVTRAD module is supported over a local Controller Area Network
by an application built using the EPICS toolkit. The EPICS database provides
controls, monitoring processes, and client connections over TCP/IP for
displays and archiving. The data retrieved from the SVTRAD module is
processed by a set of custom EPICS records and subroutines
The ADC samples are combined with temperature measurements from
other modules in the common EPICS environment to form average
dose rate measurements.

\begin{figure}[htbp]
\begin{center}
\includegraphics[width=2.25in]{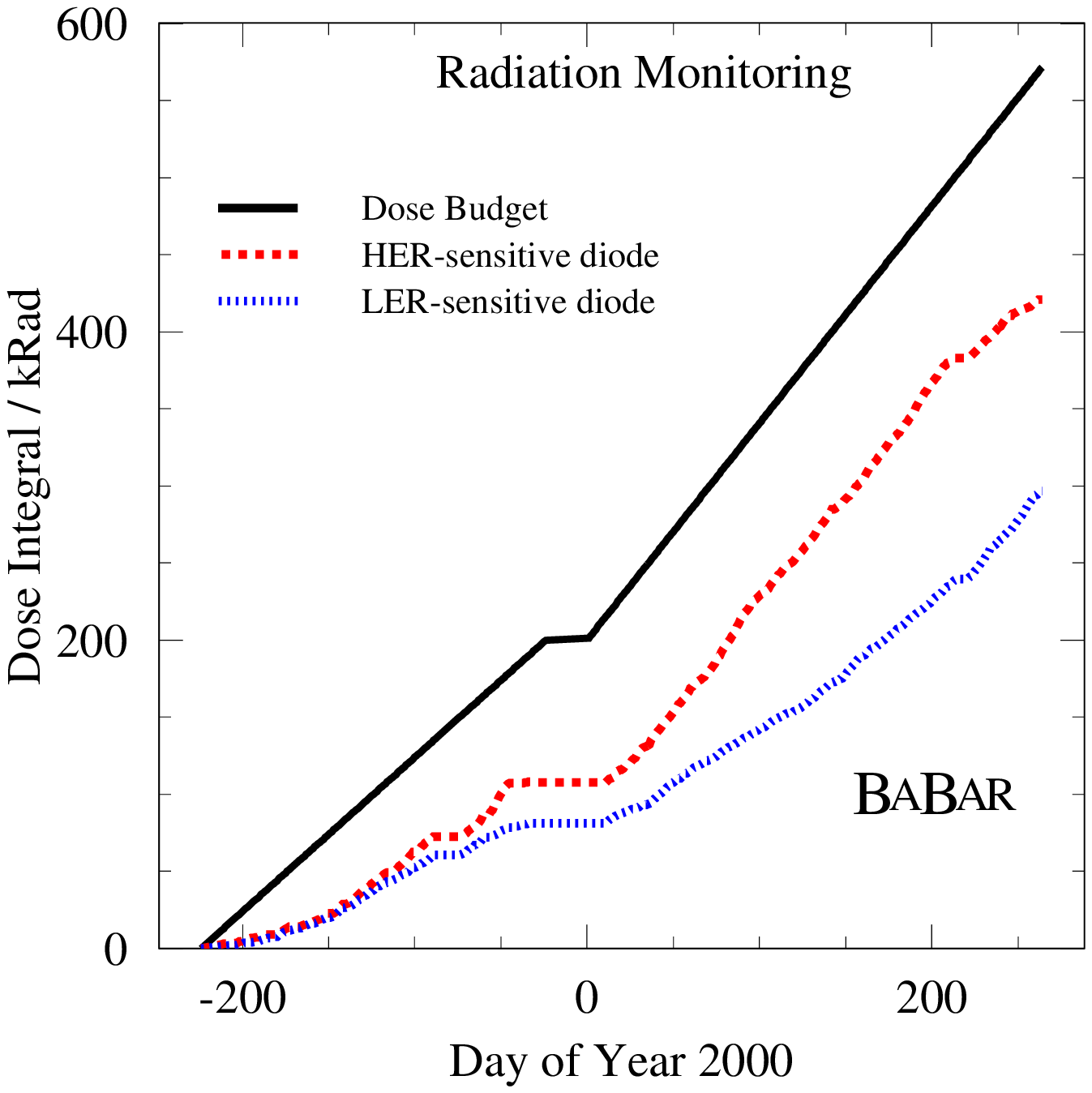} 
\includegraphics[width=2.25in]{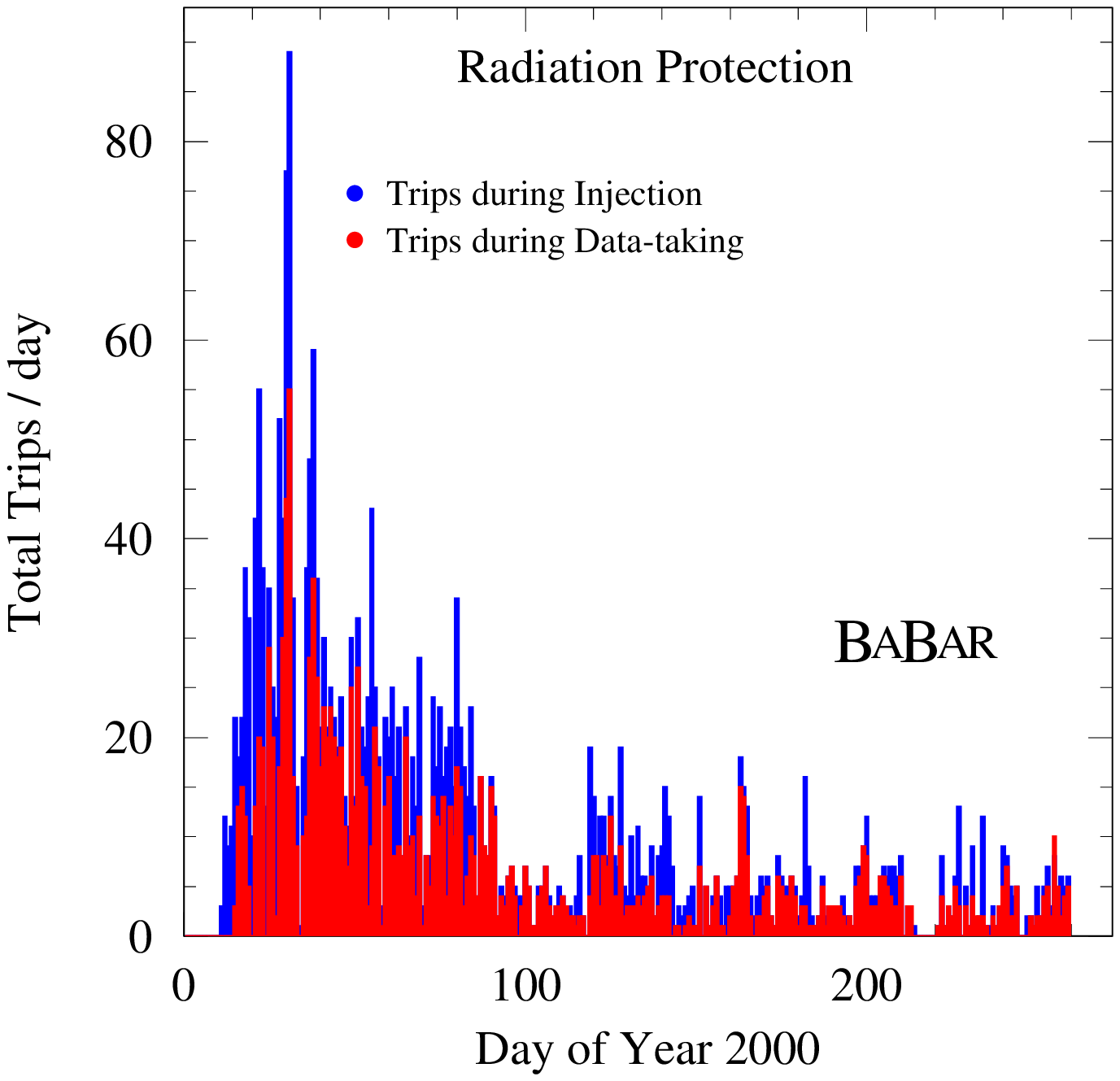}
\end{center}
\caption{(a) Integrated dose history for several
of the diode sensors, compared to the 1999-2000 SVT radiation budget.
(b) By actively aborting beams when radiation exceeds thresholds,
the SVTRAD system has successfully protected the SVT.
}
\label{fig:plots}
\end{figure}

Performance of the SVTRAD system is best illustrated in
Figure~\ref{fig:plots}(a), showing the evolution of integrated
doses for several of the diodes. In 1999 alone, there were over
4500 beam aborts caused by SVTRAD, and another 2500 in the
year 2000 (see Figure~\ref{fig:plots}(b)).
Reviewing the fast history buffer after a trip
has been an invaluable tool, providing validation of the protection
thresholds.

Systematic errors from pedestal imprecision contribute less than 10\%
absolute error to the absorbed dose integrals, and instantaneous
dose rates are measured with better than 0.25 mR/s accuracy in the
monitoring circuit, and 5 mR/s for the protection diodes. Trip thresholds
are accurate to 10 mR/s. These performance values will degrade as
integrated radiation damage exceeds 1 MRad, however.

The radiation budget of the SVT must be measured against the
constraints it places on general data-taking. The trip thresholds were
optimized for minimizing overall radiation exposure to the SVT while 
guaranteeing headroom for the accelerator to reach their performance
goals. 

The SVTRAD system has proven effective in limiting radiation exposure 
of the SVT. The online dose rate information is used extensively by 
background remediation experts to identify background sources and 
locations$^3$.

%%%%% Acknowledgments
%%
\subsection*{Acknowledgments}

We thank the PEP-II accelerator physicists and operators for 
many valuable discussions.

%%%%% Bibliography

%%%%% End Bibliography

\end{document}